\documentclass[aps,preprint,floatfix,nofootinbib,showpacs]{revtex4-1}
\pdfoutput=1
\usepackage{graphicx,color}
\usepackage{hyperref,epstopdf}
\begin{document}

\title{Various Models Mimicking the SM Higgs Boson
\footnote{Invited review in Int. J. Mod. Phys. A{\bf 27}, 1230030 (2012),
based on the plenary talk given at the 20th 
International Conference on Supersymmetry and Unification of 
Fundamental Interactions (SUSY 2012), Peking University, Beijing, China
  August 13, 2012 - August 18, 2012}
}

\renewcommand{\thefootnote}{\arabic{footnote}}

\author{
Jung Chang$^1$, Kingman Cheung$^{1,2}$, Po-Yan Tseng$^1$, 
and Tzu-Chiang Yuan$^3$}
\affiliation{
$^1$ Department of Physics, National Tsing Hua University,
Hsinchu 300, Taiwan \\
$^2$ Division of Quantum Phases and Devices, School of Physics, 
Konkuk University, Seoul 143-701, Republic of Korea \\
$^3$ Institute of Physics, Academia Sinica, Nangang, Taipei 11529, Taiwan 
}
\date{\today}

\begin{abstract}
  This review is based on the talk presented at the SUSY 2012 (Beijing).
  The new particle around 125 GeV observed at the Large Hadron Collider (LHC) 
  is almost consistent with the standard model Higgs boson, except that the diphoton decay mode
  may be excessive. We summarize a number of possibilities. 
  While at the LHC the dominant production
  mechanism for the Higgs boson of the standard model and some other 
  extensions is via the gluon fusion process, the
  alternative vector-boson fusion is more sensitive to electroweak
  symmetry breaking.
  Using the well known dijet-tagging technique to single
  out the vector-boson fusion mechanism, we investigate potential of 
  vector-boson fusion to discriminate a number of models suggested to give an enhanced 
  inclusive diphoton production rate.
\end{abstract}

\maketitle

\section{Introduction}
In this talk, we are going to summarize a few models that have been 
suggested to explain the newly observed particle of about 125 GeV 
at the Large Hadron Collider (LHC) \cite{atlas,cms}. 
It is of very high expectation that the 
observed particle is the long-sought Higgs boson, which was proposed
in 1960s \cite{higgs}. 

Before the LHC era there have been many speculations of the breaking
of electroweak symmetry (EWSB).  There are two known scales in
particle physics -- the electroweak scale and the Planck scale.
  The fundamental
Higgs boson of order 100 GeV is unstable against the radiative
corrections. The so-called gauge hierarchy problem requires new
physics has to come in around TeV scale in order that the unnatural
cancellation between the bare mass term and the higher-order terms of
the Higgs boson mass is under control. Historically, there are two
categories of models: one with a strongly-coupled EWSB sector and one
with a weakly-coupled EWSB sector. The most studied model for strong
EWSB is the technicolor-type model \cite{tc}
while that for weakly-coupled EWSB
model is the supersymmetry \cite{haber}.  
In technicolor models, the standard model
(SM) is simply an effective model below TeV scale, at which the theory
is replaced by another strong dynamics. Therefore, the cutoff scale now
becomes just TeV.  On the other hand, supersymmetric models predict
another set of particles, which differ from their SM counter parts by
a half-integral spin.  Quadratically divergent contributions
to the Higgs boson radiative
corrections are cancelled among the SM particles and the corresponding
supersymmetric particles.

If the Superconducting Super Collider (SSC) were built 
the Higgs boson could have been discovered in
early 2000s with perhaps a short running of the machine.  Since the
Higgs boson has been hiding for such a long time, many interesting
alternatives were proposed in the last 10--15 years.  Around the turn
of the century extra dimension models became very popular, not to
mention there have been a large number of varieties-- large extra
dimension \cite{add}, universal extra dimension \cite{ued}, 
Randall-Sundrum models \cite{rs},
etc. There are also the little Higgs type models \cite{lh}. In contrast to
supersymmetry the new particles have the spin as their SM counter
parts.  Contributions to the Higgs boson radiative corrections are
cancelled among the SM particles and the corresponding new
particles. Perhaps, more and more models would have been proposed if
the Higgs boson kept hiding.  Wouldn't it be more fascinating for
theorists?

The past few years the high energy community has been very excited with
a number of experimental anomalies from the Tevatron, LHC, and dark
matter (DM) experiments.
The long-time inconsistency among the DAMA results \cite{dama}
(also the CoGeNT \cite{cogent}) and
the other direct detection experiments has motivated a number of unconventional
dark matter models \cite{cdms,xenon}, 
such as inelastic DM \cite{indm}, isospin-violating DM \cite{isodm}, 
multi-component DM \cite{mdm}, etc.  
The top-quark forward-backward asymmetry observed by the CDF
and D\O\ collaborations is rather puzzling too \cite{tfb}. Many models such as
flavor-changing $Z'$, unusual $W'$, axigluon, etc \cite{tfb-t} 
were proposed, but
the new LHC results seem ruled out almost all of these models
\cite{cms-tfb}.
The CDF $Wjj$ anomaly in 2011 \cite{cdf-wjj}
also stimulated a large number of theoretical
or phenomenological models to account for the observation \cite{z-wjj}.  
However, 
with non-observation of the resonance from D\O\ and CMS \cite{d0-wjj}
the excitement gradually died out.

At the end of 2011, both the ATLAS and CMS \cite{cms-atlas} 
experiments at the LHC
have seen some excess of events of a
possible Higgs candidate in the decays of $h\to \gamma\gamma$, $h\to
WW^* \to \ell \nu \ell \nu$ and $h \to Z Z^* \to 4\ell$ channels. 
Finally, the discovery was jointly announced in July 2012 by ATLAS \cite{atlas}
and CMS \cite{cms}.  All the observed channels, $WW$, $ZZ$ and 
$\gamma\gamma$ are consistent with the predictions of the SM Higgs boson,
except that the $\gamma\gamma$ rate is somewhat higher than expectation.
The $b\bar b$ and $\tau\tau$ channels are not confirmed yet, because
of large uncertainties. 

The diphoton production rate is
about a factor of $1.3-2$ higher than that of the standard model
Higgs boson, while the $ZZ^*$ and $WW^*$ rates are consistent with the SM 
Higgs boson within uncertainties.
Nevertheless, the observed rates are consistent with
either the SM Higgs boson or some other Higgs
models. A large number of models have been put forward to account for the
observed particle at 125 GeV, including the SM, MSSM, NMSSM, UMSSM and
other MSSM-extended models, fermiophobic Higgs, 2HDM, RS radion, 
inert-Higgs doublet, triplet Higgs models, etc.  

The $H\to \gamma\gamma$ events collected by CMS and ATLAS 
can be divided into two categories: {\it inclusive} $\gamma\gamma X$ and
{\it exclusive} $\gamma\gamma j j$ (though both experiments have more refined 
sub-divisions among various classes of events).  Presumably, the inclusive  
$\gamma\gamma X$ events include all production channels such as
gluon fusion, vector-boson fusion, associated production, etc, among which 
gluon fusion dominates for production of the SM Higgs boson and most of 
the models considered in this talk, except for the fermiophobic Higgs boson.
On the other hand, exclusive $\gamma\gamma j j$ events mainly come from
vector-boson fusion and associated production, which can be further 
disentangled by jet-tagging techniques.  The vector-boson fusion produces
energetic forward jets while associated production with a $W$ or $Z$
produces jets with 
$m_{jj} \approx m_W$  to $m_Z$.
The current evidence of the Higgs boson in the diphoton channel comes 
mainly from inclusive $\gamma\gamma X$ events, simply because the inclusive
event rate is much higher than the exclusive $\gamma\gamma jj$ event rate.

The main goal of this talk is to summarize all the models
and find the parameter space of each model that 
have been proposed to explain the excess in the inclusive Higgs diphoton
events, and attempt in distinguishing the models
using the exclusive $\gamma\gamma jj$ channel 
in vector-boson fusion (VBF) \cite{ours}. 
The exclusive VBF events with $\gamma\gamma jj$
in the final state are selected using the forward jet-tagging techniques
which will be explained shortly. 
We will first choose the parameter-space region of each model that can account
for the excess in the inclusive diphoton rate, and then in that region
of parameter space we 
calculate the exclusive $jj\gamma\gamma$ VBF production rates.
We found that the exclusive $\gamma\gamma jj$ production rate in VBF 
channel can give more information to help in distinguishing
a number of popular models.

We summarize a number of models that have been used to 
account for the excess in the
inclusive $\gamma\gamma X$ data as follows.
\begin{enumerate}
\item
The SM Higgs boson \cite{higgs} is still believed to be the most 
desirable candidate.  It is still consistent with the data within uncertainty.

\item 
The lighter Higgs boson of the minimal supersymmetric standard model
(MSSM) can acquire a large radiative correction from the top-stop 
sector to achieve a mass of 125 GeV, though it has been shown 
rather difficult to achieve an enhanced diphoton rate \cite{carena,mssm}.
However, it is possible when one of the staus is light enough, just 
above the LEP limit, and so the diphoton branching ratio is enhanced 
\cite{carena}.

\item One of CP-even Higgs bosons in the next-to-minimal
  supersymmetric standard model (NMSSM) can account for the observed
  125 GeV boson with an enhanced diphoton rate \cite{nmssm}.  It could
  be the lightest or the second lightest one.  The $U(1)$-extended
  MSSM (UMSSM) \cite{umssm} and other extensions \cite{omssm} are 
  also possible to account for the observed boson.
  The analyses for these extended MSSM models are much
  more involved and deserve dedicated studies.

\item 
The lighter CP-even Higgs boson of various types of
the two-Higgs-doublet models (2HDM) \cite{2hdm}, which has enough free 
parameters in the model that allows one to achieve a large branching ratio into
$\gamma\gamma$. 

\item In the fermiophobic (FP) Higgs boson model, the Higgs boson is
  only responsible to generate the masses to $W$ and $Z$ bosons while
  the fermion masses are generated by some other means.  Since the FP
  Higgs boson does not couple to the quarks, it cannot be produced via
  gluon fusion at hadronic colliders, but only through the
  VBF and the associated production with a
  vector-boson. Nevertheless, the FP Higgs boson lighter than 130 GeV has a
  much larger branching ratio into diphoton, such that it can still
  account for the observed 
  inclusive diphoton rate at the LHC \cite{fp}.

\item In Ref.~\cite{cy}, it was pointed out that the
  Randall-Sundrum (RS) radion, with enhanced couplings to $gg$ and
  $\gamma\gamma$ due to trace anomaly, can explain the 
  excess in the inclusive diphoton production rate and 
  suppressed $WW$ and $ZZ$ rates, which provides the
  most economical alternative solution to explain the observed data.

\item
The inert-Higgs-doublet model (IHDM) \cite{arhrib}, which is a special case
of 2HDM, in which one of the
doublets entirely decouples from the leptons, quarks, and gauge bosons
while the other one takes on the role of the SM Higgs doublet.
The production rate of the Higgs boson is the same as the SM one. 
However, the decay width of $h \to \gamma\gamma$ can be enhanced by
the presence of the charged Higgs boson in the loop.  It was shown
\cite{arhrib} that the diphoton production rate can be enhanced by
 a factor of about $1 - 2$.

\item There may also be some possibilities that the SM-like 
Higgs boson first decays into two light scalar or pseudoscalar 
bosons, followed by subsequent decays into collimated pairs of 
photons, which appear as two photons in the final state \cite{draper}.
On the other hand, instead of top-down approaches,
it would also be useful to reversely determine 
the couplings and the nature of the observed 125 GeV particle by
studying all the available data \cite{dean}.

\end{enumerate}

The disadvantage of gluon fusion is that it is not clear what particles
and their masses running in the triangle loop. In some models, the 
contribution from a particular charged particle can increase or decrease
the diphoton decay width, depending on the relative signs. For example,
in the supersymmetric models, there are additional sfermions, charginos,
charged Higgs bosons running in the loop, and therefore resulting in
complicated dependence on the model parameters. 
On the other hand, the advantage of using $WW$ fusion or associated 
production with a $W$ or a $Z$ boson is that the production diagram is clean
and directly testing the couplings of $hWW$ and $hZZ$.
Furthermore, the $WW$ fusion has a cross section at least a factor of 2
larger than the associated production.  We therefore focus on $WW$ fusion.
The $WW$ fusion can be extracted by the presence of 
two energetic forward jets.  We can impose selection cuts to 
select jets in forward rapidity and high energy region \cite{barger,dieter}.
By combining the production rates in the inclusive $\gamma\gamma X$
and exclusive $\gamma\gamma jj$ channels, one can obtain useful information
about the nature of the 
125 GeV new particle recently observed at the LHC. 

We calculate the event rates in the $WW$ fusion channel 
for a number of models that have been used to interpret the current 
LHC data of the 125 GeV ``Higgs boson''.
The theoretical cleanliness of $WW$ fusion 
has been explained in the last paragraph. We believe that the $WW$ 
fusion channel can provide useful information
 to discriminate various models. The organization of this paper 
is as follows. We briefly highlight a number of 
models in the next section, and the $WW$ fusion and selection cuts in
Sec. III.  We give the decay branching ratios in Sec. IV and
production rates in Sec. V.  We conclude in Sec. VI.
 
\section{Models}

\subsection{Standard Model Higgs Boson}

The SM Higgs boson \cite{higgs} 
is still the most favorable candidate to interpret
the observed boson, though the experimental data showed slightly excess
in inclusive $\gamma\gamma X $ events over the prediction 
of the SM Higgs boson \cite{atlas,cms}. Production of the SM Higgs boson 
is dominated by gluon fusion, which is an order of magnitude larger than
the next important mechanism -- VBF.

\subsection{Two-Higgs-Doublet Model (2HDM)}
There are two Higgs doublets instead of just one in the 2HDM. 
In order to avoid dangerous tree level flavor-changing neutral currents, 
the popular 2HDMs are imposed a discrete symmetry. 
In the type I, all of the fermions couple to a single Higgs 
doublet, and do not couple to the second doublet; while in the type II, 
one doublet couples only to down-type quarks and another doublet couples 
to the up-type quarks.  
In this talk, we focus on the type II, which has the same Higgs sector 
as the MSSM. The Higgs sector consists of two Higgs doublets
\[
 H_u = \left( \begin{array}{c}
             H_u^+ \\
             H_u^0 \end{array} \right )\;, \qquad
 H_d = \left( \begin{array}{c}
             H_d^+ \\
             H_d^0 \end{array} \right )\;,
\]
where the subscripts $u,d$ denote the right-handed quark singlet field
that the Higgs doublet couples to. The electroweak symmetry is broken
when the Higgs doublet fields develop the following VEVs:
\[
\langle H_u \rangle  = \left( \begin{array}{c}
             0 \\
             v_u \end{array} \right )\;, \qquad
\langle H_d \rangle = \left( \begin{array}{c}
             0 \\
               v_d    \end{array} \right )\;.
\]
Physically,
there are two CP-even, one CP-odd, and a pair of charged Higgs bosons
after electroweak symmetry breaking (EWSB), and the $W$ and $Z$ bosons 
as well as the SM fermions, except for neutrinos, acquire masses.
The Yukawa couplings and masses for fermions can be obtained from the
following Yukawa interactions after EWSB
\[
 {\cal L}_{\rm Yuk} = -y_u \overline{Q_L} u_R \tilde H_u - 
   y_d \overline{Q_L} d_R H_d             + {\rm h.c.}
\]
where ${\tilde H}_u = i \tau_2 H^*_u$.
The parameters of the model in the CP-conserving case include
\[
  m_h,\; m_H,\; m_A, \; m_{H^+},\; \tan\beta \equiv \frac{v_u}{v_d},\;
 \alpha
\]
where $\alpha$ is the mixing angle between the two CP-even Higgs bosons.
There are enough free parameters in the Higgs potential such that
all the above parameters are free inputs to the model, in contrast to
the MSSM where the Higgs potential is highly restricted
by supersymmetry in addition to gauge symmetry.

The couplings of the two lighter and heavier CP-even Higgs bosons $h$ and $H$ 
respectively
and the CP-odd Higgs boson $A$ to
the top, bottom quarks, and taus are given by,
with a common factor of  $-i g m_f/2m_{W}$ being suppressed,
\[
  \begin{tabular}{cccc}
   &  $t \bar t$ &  $b \bar b$ & $\tau^- \tau^+$ \\
$h$: \quad & $\; {\cos\alpha/\sin\beta}\;\;  $ & $\; \; {- \sin\alpha/\cos\beta}
\;\;  $ & $\; \; {- \sin\alpha/\cos\beta}\;\;  $\\
$H$: \quad & ${\sin\alpha/\sin\beta}$ &  $ {\cos\alpha/\cos\beta}$ &  
$ {\cos\alpha/\cos\beta}$\\
$A$: \quad & $-i \cot\beta \,\gamma_5$ & $-i \tan\beta  \,\gamma_5$ & 
 $-i \tan\beta  \,\gamma_5$ 
  \end{tabular}
\]
while the charged Higgs $H^-$ couples to $t$ and $\bar{b}$  via
\[
\bar b t H^- \,: \;\;\;\; \frac{ig}{2\sqrt{2} m_{W}}\, \left[
  m_t \cot\beta \;(1+ \gamma_5) + m_b \tan\beta \;(1-\gamma_5) \right ] \;.
\]
Other relevant couplings in $WW$ fusion are those to gauge bosons
are given by,
\begin{eqnarray}
h W^+ W^- \, &:\quad& ig\; m_{W} \sin(\beta -\alpha)\; g^{\mu\nu} \; ,\nonumber\\
h Z Z \, &:\quad& ig\; m_{Z} \frac{\sin(\beta -\alpha)}{\cos\theta_{W}}\; 
  g^{\mu\nu} \nonumber \;.
\end{eqnarray}

Dominant production of the light CP-even Higgs boson $h$
at the LHC is via gluon fusion, similar to the
SM Higgs boson with the top quark running in the loop; however, in
the large $\tan\beta$ region the bottom-quark contribution can also
be substantial. 
Since the bottom-Yukawa coupling can be substantially enhanced, the
gluon fusion cross section can be larger than the SM. 
On the other hand, since the couplings of the $h$ to the $WW$ and $ZZ$ 
are simply the SM values multiplied by $\sin(\alpha-\beta)$, 
$WW$ fusion cross sections are in general smaller than the SM.

The decay into two photons is somewhat more complicated than the SM.
Besides the couplings $hWW$, $ht\bar t$, and $hb \bar b$ are different, 
there are
also the charged Higgs bosons running in the loop. The charged Higgs
boson couples to the light CP-even Higgs with the coupling \cite{mssm}
\begin{equation}
\lambda_{hH^+H^-}= \frac{m^2_h-\lambda_5v^2}{m^2_W} \cos(\beta+\alpha)+
\frac{2m^2_{H^{\pm}}-m^2_h}{2m^2_W}\sin(2\beta) \sin(\beta-\alpha) \;.
\end{equation}
However, the $b\to s\gamma$ and B meson mixing constraints require 
the charged Higgs boson mass $m_{H^\pm} > 500$ GeV for intermediate 
to large values of $\tan\beta$ \cite{otto}.  We will choose
$m_{H^\pm} = 500$ GeV in our analysis below.

The overall 
diphoton production rate $\sigma(gg\to h) \times B(h\to \gamma\gamma)$
in gluon fusion can easily vary between $0.5 -2$ of the SM prediction
depending on parameters \cite{2hdm}.
It was shown in Ref.~\cite{2hdm} that the enhancement in branching 
ratio can be obtained roughly along $\sin\alpha$ near zero for 
all $\tan\beta$ in the type II model. We choose parameter space points there to
illustrate.

\subsection{Supersymmetric Higgs boson: MSSM}

In order to achieve a mass of 125 GeV for the lighter CP-even Higgs boson,
a very large radiative correction is needed, which essentially comes
from top-stop loop.  The approximate formula for the lighter CP-even
Higgs boson is given by \cite{carena}
\begin{equation}
m_h^2 \approx m_Z^2 \cos^2 2\beta + \frac{3 m_t^4}{4 \pi^2 v^2} \left[
  \frac{1}{2} X_t + t + \frac{1}{16\pi^2} \left(
  \frac{3 m_t^2}{2 v^2} - 32 \pi \alpha_s \right ) t \left( X_t  + t \right ) 
  \right ]
\end{equation}
where 
\begin{equation}
X_t = \frac{2 (A_t - \mu \cot\beta)^2 }{M_{\rm SUSY}^2} \left( 1 - 
 \frac{ (A_t - \mu \cot\beta)^2}{12 M^2_{\rm SUSY} } \right ) \;, 
\;\;\; t = \frac{M_{\rm SUSY}^2 }{ m_t^2 }
\end{equation}
and $M_{\rm SUSY} \sim 1$ TeV is the SUSY scale. A large $A_t$ is needed
to generate a large correction.  Here we follow the findings in 
Ref.~\cite{carena} for the parameter space:
we choose $m_{Q_3} = m_{U_3} = 850$ GeV, $A_t = 1.4$ TeV, $m_A = 1$ TeV, 
and $\tan\beta=60$.  
A detailed analysis of the MSSM parameter space based on Bayesian 
statistical analysis in light of the new observation of the 125 GeV 
Higgs candidate 
is also presented recently in Ref.~\cite{mssm-bayesian}.
The reason behind such a large $\tan\beta$ is the stau
contribution to the diphoton branching ratio explained below \cite{carena}.

In the production part via gluon fusion, the difference between
the SM and supersymmetric models is that squarks
also run in the triangle loop.  As the experimental data have
pushed the squark masses
of the first two generations to be quite heavy
but not the third generation (stop and sbottom) the change in 
production rates could be substantial, especially in large $\tan\beta$.
On the other hand, the decay into diphoton is more involved
in SUSY models.  All charged particles, including
squarks, sleptons, charginos, charged Higgs boson can flow in the 
triangle loop.  With the present constraints from experiments, the
production rate into diphoton (equal to production cross section 
times the branching ratio into diphoton) in the MSSM is shown to be 
very similar to the SM one and that the diphoton production rate 
can hardly be enhanced by more than a factor of $1.5$ 
\cite{carena,mssm,mssm-bayesian}.

The formulas for the decay of the Higgs boson into two photons
as well as production via gluon fusion can be found in Ref.~\cite{hunter}.  
The couplings of the lighter CP-even 
Higgs boson to the $WW$ or $ZZ$ are given by
the SM ones multiplied by $\sin(\alpha - \beta)$. Therefore, the 
production rate in the $WW$ fusion is in general similar to or 
smaller than the SM prediction.

We look at the parameter space in which the diphoton 
production rate would be larger than the SM value in the MSSM. It was 
shown in Ref.~\cite{carena} that diphoton production rate can be larger
than the SM one if one pushes the stau to be very light, just above the
LEP limit. In addition to the above mentioned soft parameters, the
other parameters are $m_{L_3}$, $m_{E_3}$, and the $\mu$.
Without loss of generality we choose \cite{carena}
\begin{equation}
\label{stau}
 m_{L_3} = m_{e_3} = 200 - 450 \;{\rm GeV} \qquad {\rm and } \qquad
 \mu = 200 - 1000 \; {\rm GeV} \;,
\end{equation}
in which we can scan for the diphoton production rate 
$\sigma(gg\to h) B(h\to \gamma\gamma)$ to be larger than the SM rate.
The region essentially gives a light
stau, which can enhance the $B(h \to \gamma\gamma)$.  
We will scan the region 
according to Eq.~(\ref{stau}) and require the mass of the lighter 
CP-even Higgs boson around 125 GeV and the diphoton production rate larger
than the SM value.
\footnote
{There is another possibility that the heavier CP-even Higgs boson can be
at around 125 GeV and its diphoton production rates can be enhanced 
relative to the SM one \cite{js}. We will not pursue this further here.
}

\subsection{Fermiophobic Higgs}

With the name ``fermiophobic'' (FP) the Higgs boson only couples to the 
vector bosons at tree level, though higher-loop corrections can induce
small couplings to fermions. In this case, the Yukawa couplings and 
masses of fermions are generated by some other mechanisms, which are not
of concern in this talk. 

The coupling strength of the FP Higgs boson to vector bosons is 
the same as that of the SM Higgs boson.  We write the interactions as
\begin{equation}
{\cal L}_{\rm FP} = - g m_W h_{\rm FP} W^+_\mu {W^-}^{\mu}
   - \frac{g m_Z }{2\cos \theta_W} h_{\rm FP} Z_\mu Z^\mu \;.
\end{equation}
Since the FP Higgs boson does not couple to quarks, it 
cannot be produced dominantly by gluon fusion at hadronic colliders, 
but only through the vector-boson fusion and the 
associated production with a $W/Z$ boson. The corresponding 
production cross sections are the same as 
the VBF of the SM Higgs boson.
Nevertheless, the FP Higgs boson lighter than 130 GeV has a much 
larger branching ratio into diphoton, such that it can still 
account for the observed diphoton rate at the LHC \cite{fp}.
There are two reasons: (i) the FP Higgs boson decay into fermions
is highly suppressed with only the loop-induced couplings,
and (ii) the decay into photons is via a loop of $W$ boson without
the negative interference from the top quark.  Thus, the 
branching ratio into diphoton can be enhanced by more than an
order of magnitude.  Overall, the diphoton production rate at 
the LHC is comparable to the SM Higgs boson, as was used
to account for the observed boson \cite{fp}.  An earlier study
of FP Higgs boson at the LHC can be found in Ref.~\cite{andrew}.
There is basically no free parameters in this model.

\subsection{The Radion}

The RS model \cite{rs} used a warped 5D space-time, a slice of
the symmetric space $AdS_5$,
to explain the gauge hierarchy problem. 
The extra dimension $\varphi$ is a single $S^1/Z_2$ orbifold with 
one hidden and one visible 3-brane localized at $\varphi = 0$ and $\pi$,
respectively.
It was pointed out by Goldberger and Wise \cite{GW1} that the original RS model 
has a four-dimensional massless scalar (the modulus or radion) which 
does not have a 
potential and therefore the extra dimension cannot be stabilized.
A stabilization mechanism was proposed in \cite{GW1} by adding a bulk 
scalar field  propagating in the background solution which can 
generate a potential to stabilize the
modulus field.  The minimum of the potential can be arranged to give
the desired value of $k r_c \sim 12$ to solve the gauge hierarchy problem 
without extreme fine tuning of parameters.
As a consequence, the lightest excitation mode of the modulus field is 
the radion, which has a mass of
the order of 100 GeV to a TeV, and the strength of its coupling to the
SM fields is of the order of $O(1/{\rm TeV})$ \cite{GW2}.  
Phenomenology of the stabilized radion and its 
effects on the background geometry were studied in \cite{csaki}.

The interactions of the stabilized modulus (radion) $\phi$ with the SM 
particles on the visible brane 
are completely determined by 4-dimensional general covariance. Thus the radion
Lagrangian is given by
\begin{equation}
\label{T}
{\cal L}_{\rm radion} = \frac{\phi}{\Lambda_\phi} \; T^\mu_\mu ({\rm SM}) \;,
\end{equation}
where $\Lambda_\phi= \langle \phi \rangle$ is of the order of TeV and $T_\mu^\mu$
is the trace of the SM energy-momentum stress tensor, which has the 
following lowest order terms
\begin{equation}
T^\mu_\mu ({\rm SM}) =  - 2 m_W^2 W_\mu^+ W^{-\mu} 
-m_Z^2 Z_\mu Z^\mu + \sum_f m_f \bar f f + (2m_h^2 h^2 - 
\partial_\mu h \partial^\mu h  ) + \cdots \; .
\end{equation}
The coupling of the  radion to a pair of gluons (photons) is induced 
at one loop level, 
with the dominated contributions coming from the heavy top quark (top quark 
and $W$) as well as from the trace anomaly in QCD (QED). 
The expressions of the induced couplings can be found in Ref.{\cite{cy}}.
Similar to the SM Higgs boson, the dominant production channel for the
radion is via $gg$ fusion, followed by VBF \cite{coll}.
In addition, $gg\to\phi$ gets substantial enhancement from the trace anomaly.
For the decay of the radion, it is dominated by the $gg$ mode
instead of $b\bar b$ at the low mass region, while its diphoton 
branching ratio is merely a fraction of the SM value of 
$h_{\rm SM} \to \gamma \gamma$.  

Overall, the diphoton production rate $\sigma(gg \to \phi)
\times B(\phi \to \gamma\gamma)$ can be larger than the SM rate if 
the scale $\Lambda_\phi$ is small enough, and as long as it is 
consistent with the search for RS graviton.  If we do not concern about 
naturalness, the scale $\Lambda_\phi$ can be as small as 0.8 TeV.
Here we fix the scale $\Lambda_\phi$ to be $0.8-0.99$ TeV, 
which can enhance 
the diphoton production rate in gluon fusion by a factor of $1.5-1.0$ 
\cite{cy} relative to the SM rate.  Note that the branching ratios of the
radion is independent of $\Lambda_\phi$.

\subsection{Inert  Higgs  Doublet Model (IHDM)}

IHDM is a special case of 2HDM, in which one of the
doublets takes on the role of the SM Higgs doublet,
while the other one is inert which means that it entirely decouples from 
the SM leptons, quarks, and gauge bosons.
The model also has an additional $Z_2$ symmetry, for which all
SM particles are even, except for the particle content of the 
second inert Higgs doublet.
The lightest $Z_2$-odd particle of the second doublet can work as
a candidate of dark matter.
The Higgs sector consists of 
\[
  H_1 = \left(  \begin{array}{c}
                \phi_1^+ \\
                \frac{v}{\sqrt{2}} + \frac{h+ i \chi}{\sqrt{2}} 
                 \end{array}   \right ), \qquad
  H_2 = \left(  \begin{array}{c}
                \phi_2^+ \\
                \frac{S+ i A}{\sqrt{2}} 
                 \end{array}   \right ) \; .
\]
The electroweak symmetry is broken solely by one VEV:
\[
 \langle H_1 \rangle = \left(  \begin{array}{c}
                 0 \\
                \frac{v}{\sqrt{2}} 
                 \end{array}   \right ), \qquad
 \langle H_2 \rangle = \left(  \begin{array}{c}
                  0  \\
                  0
                 \end{array}   \right ) \; .
\]
The Higgs potential is given by 
\[
V = \mu^2_1 |H_1|^2 + \mu^2_2 |H_2|^2 + \lambda_1 |H_1|^4 
  + \lambda_2 |H_2|^4 + \lambda_3 |H_1|^2 |H_2|^2 
  + \lambda_4 |H_1^\dagger H_2|^2 + \frac{\lambda_5}{2}
 \left[ (H_1^\dagger H_2)^2 + {\rm h.c.} \right ]
\]
Physically, there are 2 CP-even scalars $(h, S)$, 1 CP-odd scalar $(A)$, 
and a pair of charged Higgs $(H^\pm)$.  The $h$ plays the role of the 
SM Higgs boson while the others are inert. 
A list of parameters of the model includes 
$  m_h, m_S, m_A, m_{H^\pm}, \mu_2,$ and $\lambda_2 $.

Production via gluon fusion and via $WW$ fusion are the same
as the SM Higgs boson. However, the decay into $\gamma\gamma$ 
receives additional contributions from the $H^\pm$ running in the loop.
If kinematically allowed the Higgs boson $h$ can also decay into
$ H^+ H^-,\; AA,\;$ and $S S$.  For simplicity and to achieve a large 
enough  branching
ratio into $\gamma\gamma$ we set the masses of $S, A, H^+$ to be above
the threshold $(m_h/2)$. In this model, the coupling between the 
charged Higgs boson and the SM Higgs boson $h$ is given by
\begin{equation}
g_{h H^+ H^-}=-i\frac{e}{m_W \sin \theta_W }(m^2_{H^\pm}-\mu^2_2) \;.
\end{equation}
It is clear that from this equation the sign of the charged Higgs
contribution to the triangular loop can be positive or negative,
depending on the sizes of $m_{H^+}$ and $\mu_2$. 
Thus, if we set $m_{H^+} = \mu_2$, the charged Higgs 
contribution vanishes and so the diphoton branching ratio becomes
the same as the SM one.

It was shown in Ref.~\cite{arhrib} that in gluon fusion the 
diphoton production rate is determined by the product of gluon-fusion
cross section and the $\gamma\gamma$ branching ratio
\begin{equation}
  \frac{\sigma(gg \to h) \; B(h\to \gamma\gamma)}
 {\sigma(gg \to h)_{\rm SM} \; B(h\to \gamma\gamma)_{\rm SM} } 
 = \frac{ B(h\to \gamma\gamma)}{B(h\to \gamma\gamma)_{\rm SM} } \;,
\end{equation}
which can be varied from about $0.5$ to $2$. 
The charged Higgs contribution to the diphoton branching ratio 
depends on the sizes of 
$m_{H^\pm}$ and $\mu_2$. It was shown in Ref.~\cite{arhrib} that 
when $m^2_{H^\pm} < \mu^2_2$ the diphoton branching ratio is enhanced.
Together with other theoretical constraints
one can find the region to be
$|\mu_2| \approx 100 - 200$ GeV
and $m_{H^\pm} < |\mu_2|$. 

\begin{figure}[t!]
\centering
\includegraphics[width=4in]{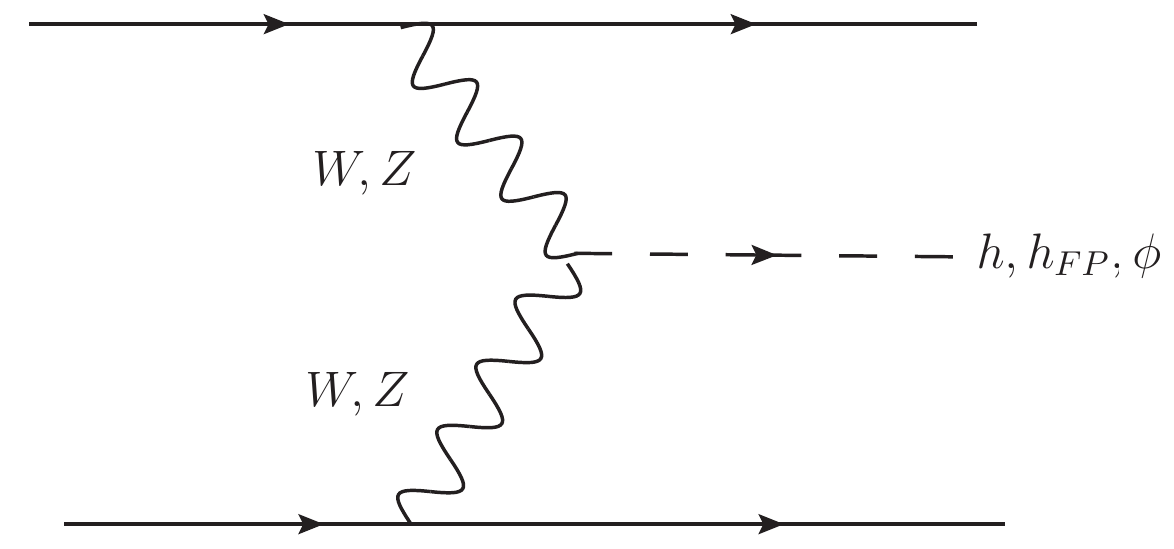}
\caption{\small \label{feyn}
A Feynman diagram showing the vector-boson fusion into SM Higgs, 
FP Higgs or radion.}
\end{figure}

\section{Vector-Boson Fusion (VBF)}

The most distinguished feature of VBF at hadronic
colliders is
the appearance of two energetic forward jets separated by a large
$\Delta R = \sqrt{ (\Delta \eta)^2 + (\Delta \phi)^2 }$, 
where $\eta$ is the pseudo-rapidity and $\phi$ is the 
azimuth angle.
The Feynman diagram is shown in Fig.~\ref{feyn}. 
Each of the initial quarks radiates
a $W/Z$ boson, which further annihilates into the Higgs boson or some
other particles under consideration. The unique feature of
this process is that the $W/Z$ bosons participating in the fusion
process are close to on-shell \cite{chan}, and irrespecive 
whether the $W/Z$ is 
longitudinal or transverse, 
the scattered quark carries almost all the energy of the incoming quark 
and goes in the forward direction."
 \cite{barger,dieter}.
This fact justifies the use of effective $W$ approximation
in all calculations for VBF in the early days.
Based on this feature we impose the following experimental cuts in 
selecting the dijet events coming dominantly from the VBF:
\begin{equation}
\label{jcut1}
E_{T_j} > 30 \; {\rm GeV},\;\;  |\eta_j| < 4.7 ,\;\; \Delta R_{jj} > 3.5 \;,
\end{equation}
and 
\begin{eqnarray}
&& \mbox{(Ejcut)} \qquad E_{j_1} > 500 \;{\rm GeV}  \qquad {\rm or} \nonumber \\
&& \mbox{(Mjjcut)} \qquad M_{jj} > 350 \;{\rm GeV} \;, \label{jcuts}
\end{eqnarray}
where the subscript ``1''  denotes the most energetic jet. 
In the cut of Eq.~(\ref{jcuts}), we choose either the energy 
of the most energetic jet $E_{j_1} > 500$ GeV or the invariant mass of the
jet pair $M_{jj} > 350$ GeV. 
This set of cuts is similar to that used by
CMS~\cite{cms-fp} and ATLAS \cite{atlas-fp} in their searches for 
FP Higgs boson.

The vector-boson fusion is well-known that it allows to probe the
direct coupling between the vector bosons and  the Higgs boson or
other particles under consideration.  This is in contrast to 
the $gg$ fusion, because any colored particles can flow in the triangular
loop and affect the production rate. For example, in MSSM
all squarks can circulate inside the loop.

\section{Decay branching ratios}

\begin{table}[th!]
\caption{\small \label{talk-tab1} 
Decay branching ratio $B(h \to \gamma\gamma)$ for the SM Higgs boson
$h_{\rm SM}$, the fermiophobic Higgs boson $h_{FP}$, the radion $\phi$,
the inert Higgs doublet model, the two Higgs doublet model, and the MSSM.
}
\begin{tabular}{lccr}
\hline
 & \multicolumn{2}{c}{Parameter Space}  & $B( h \rightarrow \gamma \gamma)$ \\
\hline
\hline
SM & &  & $2.3 \times 10^{-3}$ \\
FP & &  & $1.5 \times 10^{-2}$ \\
Radion & \multicolumn{2}{c}{$\Lambda_\phi = 0.8 - 1$ TeV} & 
$0.57 \times 10^{-3}$  \\
\hline \hline
      &  $\mu_2$ (GeV) & $m_{H^\pm}$ (GeV) &  \\
\hline 
IHDM1 &  $200$ & $70$  & $6.7 \times 10^{-3}$  \\
IHDM2 &  $200$ & $100$ & $3.3 \times 10^{-3}$  \\
IHDM3 &  $200$ & $150$ & $2.5 \times 10^{-3}$  \\
IHDM4 &  $200$ & $200$ & $2.3 \times 10^{-3}$  \\
IHDM5 &  $150$ & $70$ &  $4.2 \times 10^{-3}$  \\
IHDM10 & $100$ & $90$ & $2.4 \times 10^{-3}$  \\
\hline \hline
      & $\sin\alpha$ & $\tan\beta$ \\
\hline
2HDM1 & $0$ & $1.5$ & $3.8\times 10^{-3}$ \\
2HDM2 &  $0$ & $5$ & $6.5\times 10^{-3}$ \\
2HDM3 &  $0$ & $10$ & $6.8\times 10^{-3}$ \\
2HDM4 &  $0$ & $20$ & $6.9\times 10^{-3}$ \\
\hline
\end{tabular}
\begin{tabular}{lccccc}
\hline
\multicolumn{5}{c}{MSSM} \\
\hline
& $m_{L_3} = m_{E_3}$  & $\mu$ & $m_h$ &
  $B(h\to \gamma \gamma)$ &
  $\frac{\sigma(gg\to h) B(h\to \gamma\gamma)}
        {\sigma(gg\to h_{\rm SM}) B(h_{\rm SM} \to \gamma\gamma)}$ \\
  \hline
BP1 &250 & 400 & $127.0$ & $2.4\times 10^{-3}$ & $1.02$ \\
BP2 &250 & 500 & $126.2$ & $2.9\times 10^{-3}$ & $1.19$ \\
BP3&250 & 536 & $125.4$ & $3.6\times 10^{-3}$ & $1.45$ \\
\hline
BP4 &300 & 536 & $126.8$ & $2.4\times 10^{-3}$ & $1.005$ \\
BP5 &300 & 700 & $125.4$ & $2.8\times 10^{-3}$ & $1.15$ \\
BP6 &300 & 763 & $123.7$ & $3.4\times 10^{-3}$ & $1.38$ \\
\hline
BP7 &350 & 700 & $126.6$ & $2.4\times 10^{-3}$ & $0.999$ \\
BP8 &350 & 800 & $125.8$ & $2.5\times 10^{-3}$ & $1.03$ \\
BP9 &350 & 927 & $123.9$ & $2.7\times 10^{-3}$ & $1.11$ \\
\hline
\end{tabular}
\end{table}

Here we list the decay branching ratio into $\gamma\gamma$ for the 
SM Higgs boson, fermiophobic Higgs boson, and the radion of mass 125 GeV 
in Table \ref{talk-tab1}.
For the SM Higgs boson the decay into $\gamma\gamma$ goes
through a triangle loop of $W$ boson and top quark, between which they
interfere destructively.  
The SM Higgs diphoton branching ratio is about $2.3\times 10^{-3}$.
This is very different for the FP Higgs
boson, which allows only the $W$ boson flowing in the loop. Thus, the 
branching ratio of a FP Higgs boson into diphoton can be 
an order of magnitude larger than that of the SM Higgs boson, and also
because it does not decay into fermions.
That is the reason why it can account for the observed 125 Higgs
boson at the LHC, even though its gluon fusion production cross 
section is very small. 
The case of the radion is opposite to the FP Higgs. The diphoton 
branching ratio of the radion is a few times 
smaller than that of the SM, while its production 
rate via $gg$ fusion is substantially enhanced.

We also list the branching ratio into $\gamma\gamma$ for the Higgs boson in
the IHDM for a number of choices of 
parameters of the model in Table \ref{talk-tab1},
such that the diphoton branching ratio is enhanced relative to
the SM one: $|\mu_2| \approx 100 - 200$ GeV and $m_{H^\pm} < |\mu_2|$.
The factors affecting the
partial width into $\gamma\gamma$ are the charged Higgs boson mass and
the coupling $g_{h H^+ H^-}$. The charged Higgs loop contribution 
can interfere either
constructively or destructively with the SM contributions. 
Another
factor that would affect the branching ratio into $\gamma\gamma$ is whether
the thresholds into $SS$, $AA$, or $H^+ H^-$ are open. However, for
our choices for $m_S$, $m_{H^+}$ and $m_A$ these decays would not be allowed.
The diphoton branching ratio can be made similar to the SM one or enhanced
by a few times.

The branching ratios into $\gamma\gamma$ for the light 
CP-even Higgs boson in the 2HDM for a number of choices of 
parameters of the model are shown in Table \ref{talk-tab1}, 
such that 
the enhancement in branching ratio can be achieved roughly 
along $\sin\alpha \approx 0$. 
Similar to IHDM, the main
factors affecting the partial width into $\gamma\gamma$ are the 
charged Higgs boson mass and the coupling $g_{h H^+ H^-}$. The branching
ratio, on the other hand, also depends on other parameters such as 
$\tan\beta$ and $\sin\alpha$ as exhibited in the $hWW$ and $ht\bar t$ 
couplings. 
Along $\sin\alpha \approx 0$, the factor $\cos\alpha \approx 1$
and the factor $\sin(\beta - \alpha) \approx 1$ for large $\tan\beta$,
and thus the couplings $hWW$ and $ht\bar t$ are close to their SM values.
On the other hand, along $\sin\alpha=-1$ for large $\tan\beta$,
the $hWW$ coupling proportional to $\sin(\beta-\alpha)$ is only about 
$1/\tan\beta$. 
That is the reason why its branching ratio into $\gamma\gamma$ is very small.

\begin{figure}[th!]
\includegraphics[width=6in]{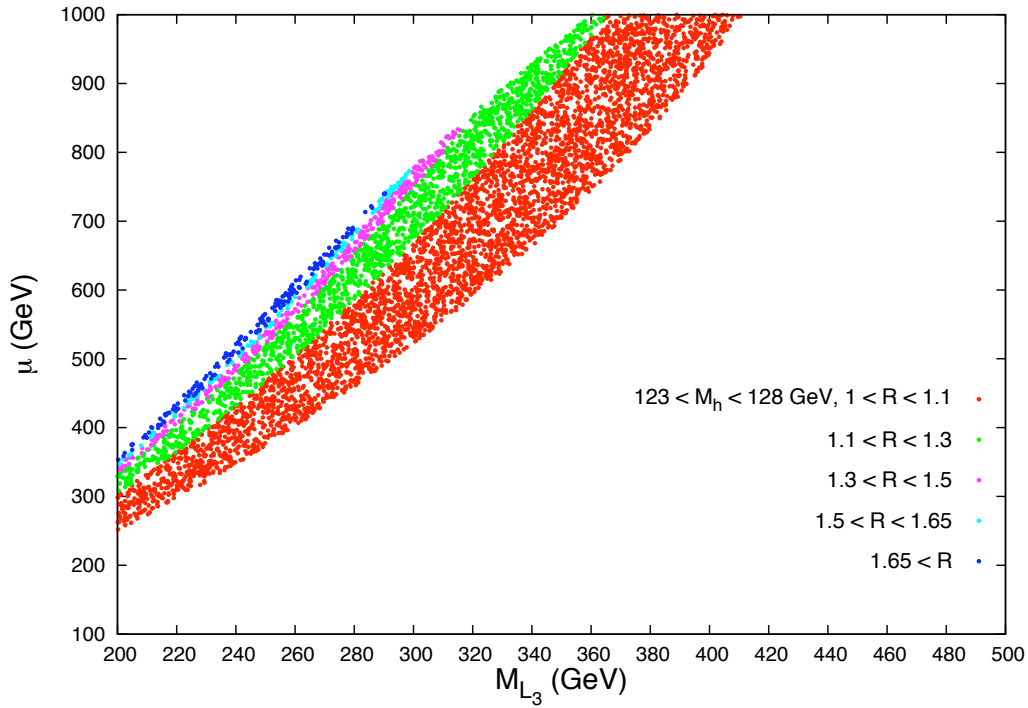}
\caption{\small \label{me-mu}
The parameter space region where  $m_h = 123 -128$ GeV and 
$R = \sigma(gg\to h) B(h\to \gamma\gamma)/
\sigma(gg\to h_{\rm SM}) B(h_{\rm SM} \to \gamma\gamma)$
is larger than 1. The other fixed parameters are
$m_{Q_3} = m_{U_3} = 850$ GeV, 
$A_t = 1.4$ TeV, $\tan\beta = 60$ and $m_A = 1$ TeV. 
}
\end{figure}

In the MSSM, we choose the region where the lighter CP-even Higgs boson is 
around 125 GeV ($123-128$ GeV) and the diphoton
production rate $\sigma(gg\to h) B(h\to \gamma\gamma)$ is equal to
or larger than the SM value. 
As explained above the stop sector
must be heavy in order to achieve a mass of 125 GeV for the lighter
CP-even Higgs boson. 
First, we fix the $m_{Q_3} = m_{U_3} = 850$ GeV and $A_t = 
1.4$ TeV with $\tan\beta = 60$ and $m_A = 1$ TeV. Second, we 
vary $m_{E_3}=m_{L_3}$ and $\mu$ to achieve a light stau so as to
enhance the diphoton production rate, according to Eq.~(\ref{stau}).
We used the FeynHiggs \cite{feynhiggs} to 
evaluate the branching ratio $B(h\to\gamma\gamma)$ and the diphoton
production rate $\sigma(gg\to h) B(h\to \gamma\gamma)$ relative to the
SM values.  We show in Fig.~\ref{me-mu} the region in the plane of 
$(m_{E_3}=m_{L_3}, \mu)$ that the $m_h = 123 -128$ GeV and the ratio
$\sigma(gg\to h) B(h\to \gamma\gamma)/
\sigma(gg\to h_{\rm SM}) B(h_{\rm SM} \to \gamma\gamma)$
is larger than 1. 
We also show the branching ratios into $\gamma\gamma$ for the lighter CP-even
Higgs boson and the inclusive diphoton production rate for a few 
selective points of MSSM  in Table~\ref{talk-tab1}.

\section{Production Rates in Vector-Boson Fusion}

\begin{table}[th!]
\caption{\small \label{talk-prod1}
Production rates in fb at PGS level
for $pp \to jj h \to jj \gamma\gamma$ at LHC-7 (LHC-8, LHC-14)
}
\begin{tabular}{lcc}
\hline
      &  photon cuts and Ejcut &   photon cuts and Mjjcut\\
\hline
SM   & $0.15\; (0.19,0.61)$ & $0.33\;(0.41, 1.1)$\\
FP   & $1.03\;(1.27,4.12)$&   $2.24\;(2.78,7.35)$ \\
Radion & $0.0038\;(0.0047,0.014)$ & $0.0076\;(0.0095,0.026)$ \\
\hline
IDHM1 &   ${0.44\;(0.56,1.79)}$&  ${0.97\;(1.21,3.23)}$ \\
IDHM2 & ${0.22\;(0.28,0.88)}$&   ${0.48\;(0.59,1.59)}$ \\
IDHM3 &  ${0.16\;(0.21,0.67)}$&   ${0.36\;(0.45,1.21)}$ \\
IDHM4 & ${0.15\;(0.19,0.62)}$ & ${0.33\;(0.41,1.11)}$ \\
\hline
IDHM5 &   ${0.28\;(0.35,1.12)}$&  ${0.61\;(0.76,2.03)}$ \\
IDHM10 & ${0.16\;(0.20,0.64)}$&   ${0.35\;(0.43,1.16)}$ \\
\hline
2HDM1 &${0.17\;(0.22,0.70)}$& ${0.38\;(0.47,1.25)}$\\
2HDM2 &${0.41\;(0.52,1.66)}$& ${0.90\;(1.11,2.99)}$ \\
2HDM3 &${0.44\;(0.56,1.79)}$& ${0.97\;(1.20,3.23)}$  \\
2HDM4 &${0.45\;(0.58,1.85)}$& ${1.00\;(1.24,3.33)}$  \\
\hline
\end{tabular}
\begin{tabular}{lccc}
\hline
MSSM BP &  photon cuts and Ejcut &  photon cuts and Mjjcut\\
\hline
BP1 & $ {0.19\;(0.28,0.83)}$ & $ {0.44\;(0.57,1.47)}$ \\
BP2 & $ {0.22\;(0.33,0.97)}$& $ {0.52\;(0.66,1.76)}$ \\
BP3  & $ {0.29\;(0.40,1.18)}$ & ${0.63\;(0.82,2.10)}$ \\
\hline
BP4  & $ {0.19\;(0.28,0.85)}$ & $ {0.43\;(0.56,1.46)}$ \\
BP5  & $ {0.22\;(0.32,0.92)}$ & $ {0.50\;(0.65,1.65)}$ \\
BP6  & $ {0.27\;(0.38,1.07)}$ & $ {0.61\;(0.75,1.90)}$ \\
\hline
BP7  & $ {0.20\;(0.26,0.85)}$ & $ {0.43\;(0.53,1.47)}$ \\
BP8  & $ {0.21\;(0.29,0.85)}$ & $ {0.44\;(0.59,1.48)}$ \\
BP9  & $ {0.22\;(0.30,0.92)}$ & $ {0.49\;(0.59,1.66)}$ \\
\hline
\end{tabular}
\end{table}

We calculate the VBF cross sections for $pp \to jj h$ for various models
under consideration. We impose the selection cuts for energetic forward
jets as in Eqs.~(\ref{jcut1}) and (\ref{jcuts}). We let the Higgs boson 
decay into $\gamma\gamma$, and impose the following cuts on the photons:
\begin{equation}
\label{gcuts}
 E_{T_\gamma} > 30\;{\rm GeV},\qquad |\eta_\gamma | < 2.5,\qquad
| m_{\gamma\gamma} - m_h | < 3.5 \;{\rm GeV}  \;.
\end{equation}
In the following, we present numerical results in both parton level
and detector-simulation level by employing the PYTHIA-PGS 
(PYTHIA v6.420, PGS4(090401)) package inside MADGRAPH \cite{mad}.

We employ MADGRAPH \cite{mad} to calculate the production cross sections
for $pp \to jj h \to jj \gamma\gamma$ and implementing the selection cuts
for the forward jets and the diphoton. The production rates of diphoton
for various models are then obtained by multiplying the corresponding 
diphoton branching ratios.

We use PYTHIA \cite{pythia} for parton showering and hadronization. 
During the parton showering we turn on the initial and final state 
QED and QCD radiations (ISR and FSR), and fragmentation/hadronization 
according to the Lund model.
We use PGS \cite{py-pgs} for detector-simulation with the most 
general settings for the LHC.
Electromagnetic and hadronic calorimeter resolutions are set at 
$0.2/ \sqrt{E}$ and $0.8/ \sqrt{E}$, respectively. 
We use the cone algorithm for jet finding with a cone size of
$\Delta R=0.5$, Sagitta resolution of $13\; \mu$m  ($\delta
p_T/p_T=1.04\times 10^{-4}$), track-finding efficiency of $0.98$,
and minimum track $p_T$ of 1 GeV. 
More details can be found in Refs.~\cite{mad,py-pgs}.

The numerical results for LHC-7, LHC-8, and
LHC-14 are listed in Table~\ref{talk-prod1} for the SM, FP, radion, IHDM, 
2HDM, and the MSSM.

It would be more transparent to show the
production rate relative to the SM one
\begin{equation}
\label{rat}
\frac{\sigma (pp \to jj X ) \times B(X \to\gamma\gamma) }
   {\sigma (pp \to jj h_{\rm SM} ) \times B(h_{\rm SM} \to\gamma\gamma) }\;,
\end{equation}
where $X$ stands for the SM Higgs or any other Higgs-like candidate 
in various models.
We found that this ratio is quite robust against various cuts (Mjjcut
or Ejcut as in Eq.~(\ref{jcuts})) and against the energy of the collision.

\begin{figure}[t!]
\centering
\includegraphics[width=6.7in]{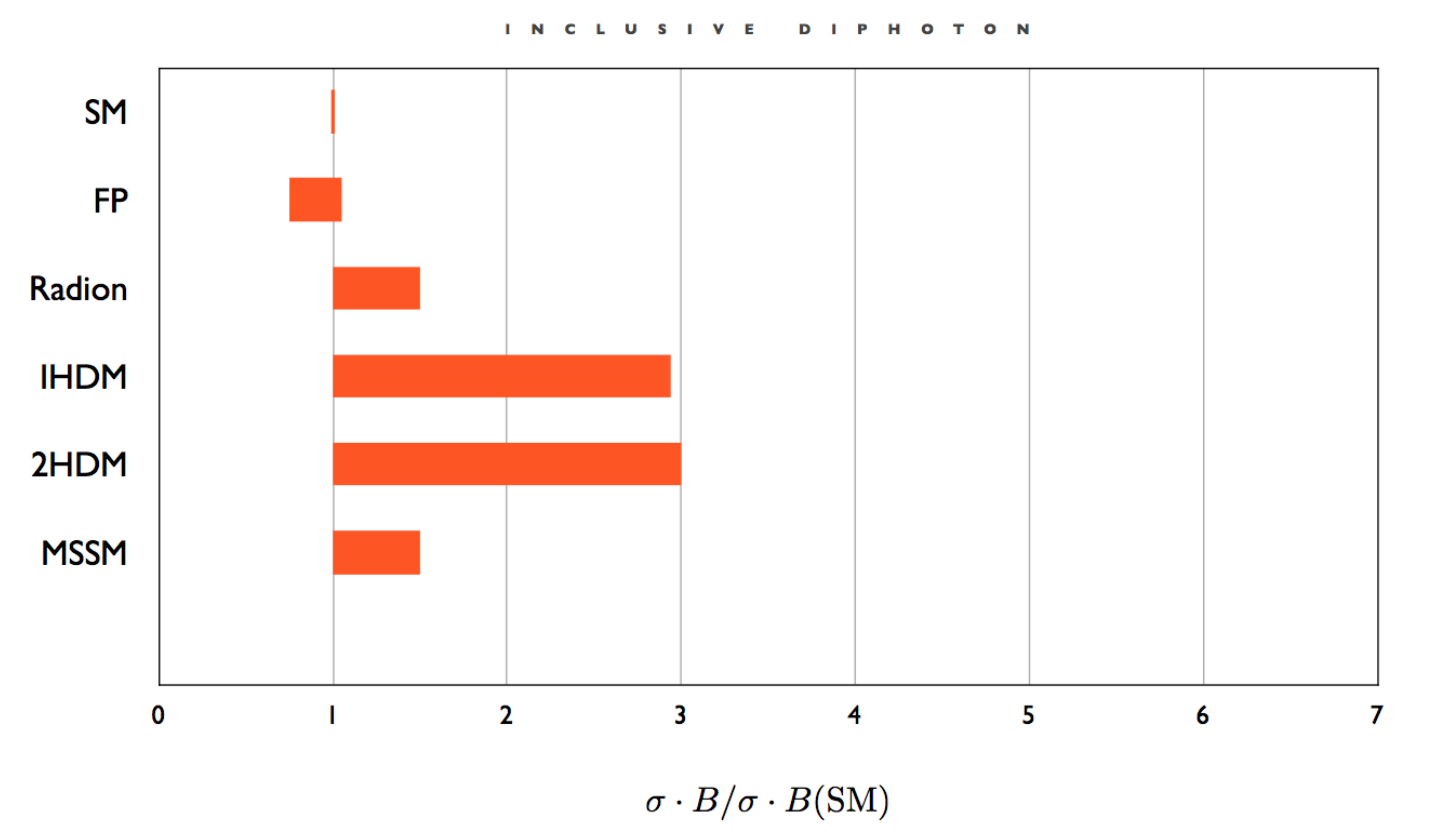}
\includegraphics[width=6.7in]{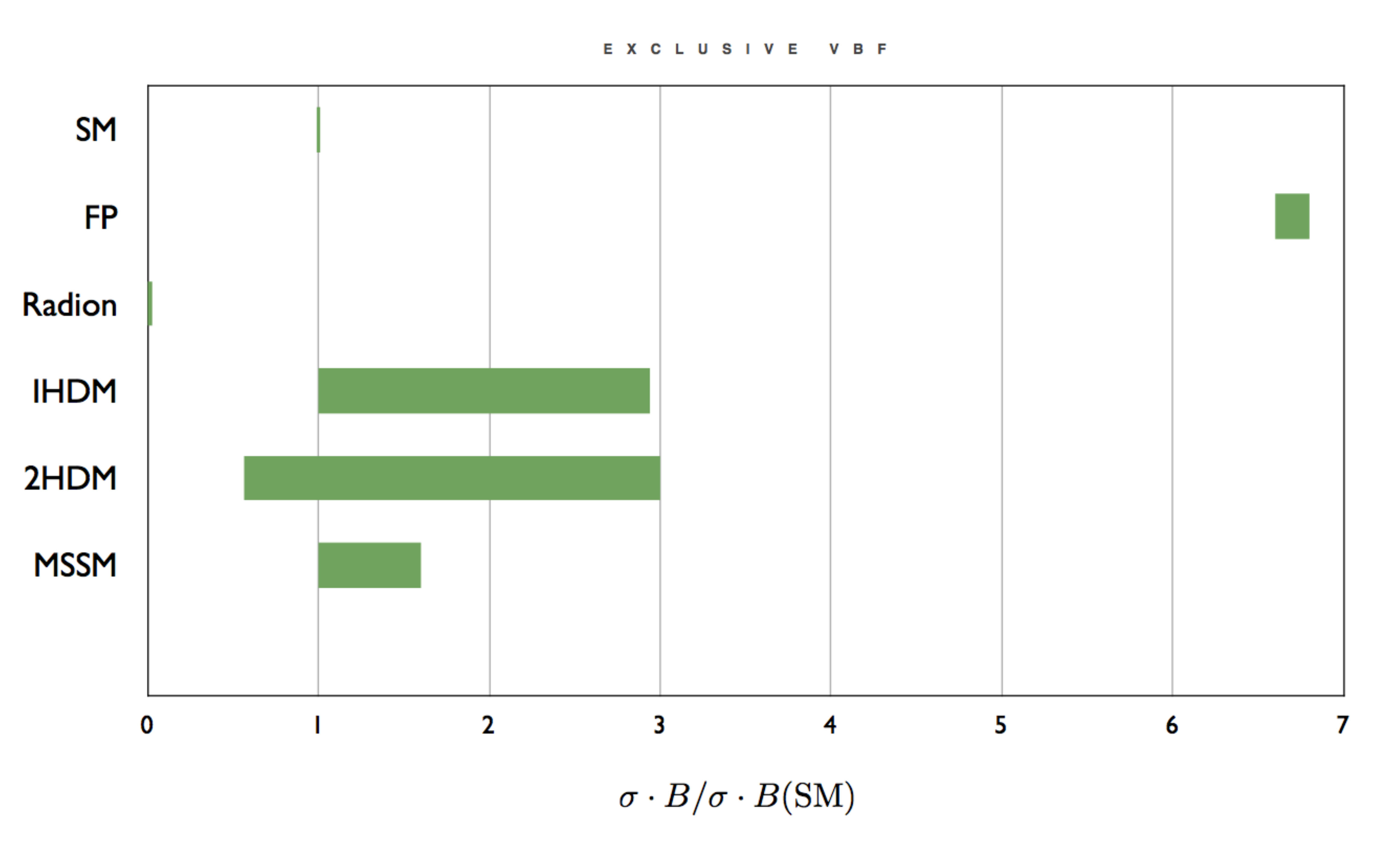}
\caption{\small \label{ratio}
{\it Upper:} the ratio of {\it inclusive} diphoton production rates
  $\frac{\sigma (X) \times B(X \to\gamma\gamma) }
      {\sigma (h_{\rm SM} ) \times B(h_{\rm SM} \to\gamma\gamma) }$ 
for various models. The parameter space is chosen such that the ratio is
equal to or larger than 1, except for the FP Higgs boson which has no free parameter.
{\it Lower:} 
the ratio of 
the {\it exclusive} $jj\gamma\gamma$ production rate
$\frac{\sigma (pp \to jj X ) \times B(X \to\gamma\gamma) }
   {\sigma (pp \to jj h_{\rm SM} ) \times B(h_{\rm SM} \to\gamma\gamma) }$ 
for each model in the corresponding parameter space.
}
\end{figure}

In the upper panel of Fig.~\ref{ratio}, we first show the ratio for 
the {\it inclusive} diphoton production rate
  $\frac{\sigma (X ) \times B(X \to\gamma\gamma) }
      {\sigma (h_{\rm SM} ) \times B(h_{\rm SM} \to\gamma\gamma) }$ 
of each model, which is dominated by gluon fusion, except for the 
FP Higgs boson.
The parameter space of each model is chosen such that this inclusive
diphoton rate is equal to or larger than the SM rate
except for the FP Higgs boson which has no free parameter.  
For the same parameter space, we show in the lower panel the ratio for 
the {\it exclusive} $jj\gamma\gamma$ production rate
$\frac{\sigma (pp \to jj X ) \times B(X \to\gamma\gamma) }
   {\sigma (pp \to jj h_{\rm SM} ) \times B(h_{\rm SM} \to\gamma\gamma) }$ 
for each model.
The figure is valid for LHC-7, LHC-8 and LHC-14, and for 
using either Mjjcut cut on both forward jets 
or Ejcut cut on the most energetic jet.

It is clear that the models can be employed to explain the
excess in the inclusive diphoton rates in some parameter space region,
but for the same region of parameter space the ratio of exclusive
VBF production would be different among the models. The FP
Higgs would be a number of times larger than the SM in the VBF channel,
but the RS radion would give negligible exclusive VBF production. 
On the other hand, the IHDM, 2HDM, and the MSSM would give similar
ratios in both inclusive and exclusive production. 
The 2HDM can give a somewhat smaller ratio in the exclusive VBF.

\subsection{Background Discussion}

The experimental details for VBF in the context of searching for 
the FP Higgs boson are given in Ref.~\cite{cms-fp}. 
The dominant backgrounds to $h \to \gamma\gamma$ consist of i)
the irreducible background from the prompt diphoton production, and ii) 
the reducible backgrounds from $pp \to \gamma + j$ and $pp \to j j$,
when one or more of the jets hadronize into (typically neutral) pions and
deposit energy in the electromagnetic calorimeter.
These reconstructed objects are generally referred to as fake photons. 
Isolation is a very useful tool to reject the non-prompt background 
coming from electromagnetic showers originating in jets,
because these fake photons are often accompanied by single and multiple 
$\pi^0$s.

The dijet tag, as defined in Eqs.~(\ref{jcut1}) and (\ref{jcuts}), 
together with the Higgs boson decaying into diphoton 
selects a special class of events consisting of two photons and two forward
energetic jets. A signal-to-background ratio of order 1 can be achieved
\cite{cms-fp}.
The signal from $h \to \gamma\gamma$ will be identified as a sharp peak
in the $m_{\gamma\gamma}$ distribution, where the background is in
continuum. In Ref.~\cite{cms-fp}, the background model is derived from
data. In the new run at 8 TeV, we expect the same treatment is applied
to the background.  
The estimation of significance of each signal is beyond the scope of
the present paper.

Given that we have obtained the event rates of each model in the
tables, which have been under experimental cuts and simulations (PGS),
it is straight-forward to estimate the required luminosity to probe
various scenarios. The continuum background at the 7 TeV and 8 TeV LHC
in VBF channel has been obtained in Refs.~\cite{cms-fp} and
\cite{cms}.  Since the SM Higgs boson has been seen above the
background with some significance level in the VBF channel in the
current LHC runs (about 5 fb$^{-1}$ at 7 TeV and {\color{red}about} 5 fb$^{-1}$ at
8 TeV), the scenarios with higher event rates than the SM Higgs boson
should be detectable.  The projected integrated luminosity at the end
of 2012 is about $10-15$ fb$^{-1}$ for each experiment, it would not
be a problem to investigate the scenarios in the VBF channel.

\section{Conclusions}

LHC is expected to confirm if there is a new particle at 125 GeV by the end
of this year. The likelihood for a new discovery is rather high. Nevertheless,
whether this new particle is the SM Higgs boson is not an easy question to 
answer. 

Here the scenario of this 125 GeV particle produced by
vector-boson fusion, instead of gluon fusion as the dominant production
mechanism for the standard model Higgs boson, is studied in details.
By using the forward dijet tagging technique, one can single out the 
vector-boson fusion mechanism. 
We studied a number of popular new physics models that have been employed to
interpret the observed particle at 125 GeV, 
including fermiophobic Higgs boson,  the Randall-Sundrum radion, 
inert-Higgs-doublet model, two-Higgs-doublet model, and the MSSM.
Since the inclusive diphoton channel showed an excess over the SM predictions,
we first selected the parameter space of each model that can give
an inclusive diphoton rate larger than or equal to the SM rate. 
Then, we calculate the exclusive $jj\gamma\gamma$ diphoton production rate
in VBF for that parameter space.
If the diphoton mode excess seen at LHC-7 can be firmly established by 
the new LHC-8 data, it will be the utmost task to identify the
nature of this particle.  Perhaps, it is simply the SM Higgs boson
with some level of statistical fluctuation, but it could also be
the RS radion \cite{cy}, fermiophobic Higgs boson \cite{fp},
the light CP-even Higgs boson of the 2HDM \cite{2hdm}, 
the Higgs boson of the IHDM \cite{arhrib}, or one of the 
CP-even Higgs bosons in other extensions of the MSSM \cite{nmssm,umssm}, 
all of which  can allow an enhancement to the $\gamma\gamma$ production 
rate.
On the other hand, the vector-boson fusion, as singled out by the dijet tag,
provides  useful information in helping to differentiate among
various models.
It is not hard by browsing through Fig.~\ref{ratio} to conclude the 
following 
\begin{itemize}
\item 
If a similar rate is seen in inclusive production but
no large excess is seen in 
the exclusive VBF production it would unlikely
be a fermiophobic Higgs boson.

\item
If a similar rate or excess is seen in inclusive production 
but some events are seen in
exclusive VBF production rate, it would unlikely be the RS radion.

\item
If moderate excess is seen in both inclusive production and exclusive VBF
production, it could be the Higgs boson of the IHDM, 2HDM, or the MSSM.
However, if the excess is over 60\% 
it will pose severe challenge to the MSSM.

\end{itemize}
It seems easy to rule out either fermiophobic Higgs boson or RS radion,
providing that we see no large excess or some events in VBF channel, 
respectively.
However, it is still difficult to distinguish the other models when
the inclusive and exclusive rates are similar to or slightly larger than
the SM values.
If the production rate of the diphoton mode at 125 GeV lines up with
the SM prediction eventually, it is still premature to conclude this
is coming from the SM Higgs boson. Other alternatives in MSSM, NMSSM,
or other SUSY models can also be mimicking the SM Higgs boson,
depending on the parameter space of the new physics model.
In any case, once the signals at 125 GeV are confirmed further
studies including all possible decay modes are to be taken into
account in order to discriminate these many alternatives beyond the
standard model.  

Vector-boson fusion is the next most important production mechanism
that must be taken into account to fully identify the newly discovered 
particle.

\section*{Acknowledgment}
This work was supported in part by the National Science Council of
Taiwan under Grants No. 99-2112-M-007-005-MY3 and No.
101-2112-M-001-005-MY3 as well as the
WCU program through the KOSEF funded by the MEST (R31-2008-000-10057-0).


\end{document}